\documentclass[a4paper,11pt]{article}
\pdfoutput=1 

\usepackage{jinstpub} 
\usepackage{subcaption} 

\usepackage{lineno}
\usepackage{booktabs, tabularx, multirow, array} 

\title{\boldmath Coating $\mu$m TPB on a cylindrical detector and studying the sample films being cooled to LN and LHe temperatures}

\author[a]{Jiangfeng Zhou,}
\author[b]{Zebang Ouyang,}
\author[a,c,1]{Junhui Liao,\note{Corresponding author.}}
\author[a]{Zhuo Liang,}
\author[a]{Zhaohua Peng,}
\author[d]{Lei Zhang,}
\author[d]{Lifeng Zhang,}
\author[a]{Jian Zheng,}

\affiliation[a]{Division of Nuclear Physics, China Institute of Atomic Energy, Sanqiang Rd. 1, Fangshan district, Beijing,\\ China, 102413.}
\affiliation[b]{School of Nuclear Technology, University of South China, ChangSheng West Rd. 28, Hengyang, Hunan,\\ China, 421009.}
\affiliation[c]{Department of Physics, Brown University, Hope St. 182, Providence, Rhode Island, USA, 02912.}
\affiliation[d]{Division of Nuclear Synthesis Technology, China Institute of Atomic Energy, Sanqiang Rd. 1, Fangshan \\ district, Beijing, China, 102413.}

\emailAdd{junhui\_liao@brown.edu}

\abstract{ALETHEIA is a newly established dark matter direct detection project that aims at hunting for low-mass WIMPs. TPB  is widely implemented in liquid helium and argon experiments to shift VUV photons to visible light. We first report that we have successfully coated $\sim 3 ~\mu$m TPB  on the inner walls of  a 10-cm cylindrical PTFE detector; we split the coating process into two steps to have all of the surfaces being coated with the same thickness; three independent methods were applied to figure out the thickness of the TPB coating layers, and consistent results were obtained. Second, with an SEM machine, we scanned the surface of TPB coating sample films exposed to different cryogenic temperatures. The first group of sample layers were immersed into a liquid nitrogen dewar for forty hours, the second group samples were cooled to 4.5 K for three hours, and the third group stayed at room temperature after coating. The SEM-scanned images of the sample films barely show any noticeable difference.}


\arxivnumber{2210.12735} 

\collaboration[c]{on behalf of ALETHEIA collaboration}

\begin{document}
\maketitle
\flushbottom

\section{Introduction to the ALETHEIA project}
\label{sec:intro}

ALETHEIA (A Liquid hElium Time projection cHambEr In dArk matter) is a low-mass DM (Dark Matter) direct detection experiment with the ROI (Research Of Interest) of $\sim$ 100s MeV/c$^2$ - 10 GeV/c$^2$. The ALETHEIA project will implement the arguably most competitive technology in the community of DM direct detection, TPC (Time Projection Chamber), and will be filled with the arguably cleanest material, LHe (Liquid Helium-4). 

In the community of DM direct detection, LAr and LXe TPC experiments, such as DarkSide~\cite{DarkSideWebsite}, DEAP~\cite{DEAP3600Website}, LUX~\cite{LUXWebsite}, LZ~\cite{LZWebsite}, PandaX~\cite{PandaXWebsite}, and XENON~\cite{XENONWebsite}, have led high-mass DM searches for more than a decade in one way or another. Among other legacies, the two most important ones of such successful experiments might be: (a) a liquid noble gas TPC can scale up from less than 1 kg to multiple tons, and (b) a series of techniques can be applied to decrease detectors' backgrounds continuously, including but not limited to: elegant detector design, detector's bulk material online purification, events' 3D reconstruction and selection, detector's fiducial volume choice, frequent detector online calibrations, and reliable discriminations for ER (Electron Recoil) and NR (Nuclear Recoil) events originated from the difference of the stopping power (dE/dx) of electrons and ions.

Helium ($^4$He) is the lightest noble element and the second lightest element, so it is more sensitive to low-mass DM than peer experiments implementing much heavier elements such as xenon, germanium, argon, and silicon since the recoil energy is roughly inversely proportional to the mass of the detector material~\footnote{Although the recoil energy of helium is roughly a factor 7 and 18 higher than silicon and germanium, respectively, the semiconductor's ionization energy is $\sim$ 3 eV, which is a factor of 8 smaller than LHe, 24.6 eV. So effectively, in terms of ionization energy, LHe does not have advantages over semiconductor detectors. However, as will be discussed below, an LHe TPC could scale up to ton-scale and achieve extremely low backgrounds.}. We estimate that an LHe TPC is sensitive to sub-GeV DM by considering helium's ionization and excitation energy.
Helium has only one isotope, $^3$He, with a natural abundance of 0.000137\% on Earth's atmosphere~\cite{heliumAbundance}. Both $^4$He and $^3$He are not radioactive, so the intrinsic background of helium is null. As a comparison, argon and xenon have significant backgrounds contributed by radioactive isotopes. Moreover, nothing is solvable in LHe~\footnote{In the paper, LHe refers to ``Helium I'', which means the temperature is between 2.17 K and $\sim$ 5 K. Helium gas becomes liquid as long as being cooled to $\sim$ 5 K; this is the ``Helium I'', or ``general'' LHe. Keep cooling to 2.17 K or below, and LHe reaches ``Helium II'' or the superfluid status.}, except $^3$He and hydrogen. As mentioned above, $^3$He abundance is as low as 1.4E-6. Moreover, for hydrogen, the solubility is as low as 10$^{-14}$~\cite{JEWELL1979682} at 1 K temperature. That is to say, impurities in LHe would show up as solid states since they are not solvable. Therefore, they can be easily purified with getters and cold traps before being filled into the TPC. 
As summarized by the ALETHEIA review panel, thanks to ``its powerful combination of intrinsically low radioactivity, ease of purification, and charge/light discrimination capability,'' the ALETHEIA detector could achieve an ``extremely low background.''~\footnote{This statement was copied from the reviewing documentation of the ALETHEIA project~\cite{ALETHEIA_2019Review}. For more details about the review, please see below.}

As mentioned above, LAr and LXe DM experiments have demonstrated that liquid noble gases TPCs can be scaled up to 10 tonnes and achieve low backgrounds. Therefore, an LHe TPC is expected to be built at the ton-scale (though unique technical challenges are foreseen). Furthermore, combining the LHe's unique advantages on backgrounds control, we expect that a ton-scale LHe TPC could reach extremely low backgrounds, therefore, fully touch down the $^8$B neutrino floor ($\sim$10$^{-45}$ cm$^2$) or even below. For more details, please refer to our submitted paper~\cite{ALETHEIA-EPJ-22}.

However, based on the authors' knowledge, no single LHe TPC has ever been built worldwide, so building the ALETHEIA detector naturally has some technological risks. To understand the challenges and possible solutions, we invited a few world-leading physicists in the communities of DM direct detection, LHe, and TPC~\footnote{The review panel members are: Prof. Rick Gaitskell at Brown University, Prof. Dan Hooper at Fermilab and the University of Chicago, Dr. Takeyaso Ito at Los Alamos National Laboratory, Prof. Jia Liu at Peking University, Prof. Dan McKinsey at UC Berkeley and Lawrence Berkeley National Laboratory, and Prof. George Seidel at Brown University.} to review the project in Oct 2019~\cite{DMWS2019PKU}. The panel evaluated the project very positively and presented insightful comments and practical suggestions. One of them is studying scintillation with small LHe cells. 

When passing through LHe, energetic particles deposit all or part of their kinematic energy; some will show up as scintillation peaked at 80 nm. However, no commercial photosensors can detect the 80 nm light directly with high efficiency. Therefore, we must convert it into visible light with a wavelength shifter. TPB (Tetraphenyl Butadiene,1, 1,4, 4-tetraphenyl-1, 3-butadiene) is a common conversion material and has been implemented in detectors at 4 K or below~\cite{McKinsey03, Ito2012, ItoSeidel13, Seidel14, Ito16, Phan20, SpiceHeRald21}. 

In the paper, we first briefly report the process of coating a few $\mu$m TPB on the inner walls of a 10-cm cylindrical PTFE cell, as shown in section~\ref{secCoating}. In section~\ref{secCoatingScanning}, we then compare the Scanning Electron Microscope (SEM) images on TPB sample films after being immersed into Liquid Nitrogen (LN), cooled to $\sim$ 4 K with a G-M cryocooler, and with no cooling. The SEM scanning images show (a) TPB has been well coated on all of the sample films, and (b) compared to the TPBs without being cooled, no noticeable difference for the ones that went through low temperatures. 

\section{TPB coating on a 10-cm size cylindrical detector}
\label{secCoating}
\subsection{TPB Coating Introduction}
\label{secCoatingSubIntroduction}

As demonstrated in references~\cite{McKinsey03, Ito2012, ItoSeidel13, Seidel14, Ito16, Phan20, SpiceHeRald21}, TPB  is capable of converting 80 nm photons into visible light in LHe (more specifically, superfluid helium). At room temperature, according to the reference~\cite{Benson18}, $\sim 3~ \mu$m is the most appropriate thickness for TPB in terms of maximizing conversion yield for 80 nm light, $\sim$ 30\%. The TPB layer in DEAP-3600~\footnote{DEAP stands for Dark matter Experiment using Argon Pulse-shape discrimination.} detector is 3 $\mu$m~\cite{Broerman2017}, though DarkSide-50 and DarkSide-20k have a much more thicker TPB layer up to $\sim 200~ \mu$m~\cite{DarkSide20k17, AGNES2015456}.

We referred to the coating process of the DEAP-3600 experiment, as introduced in references ~\cite{PollmannPhDThesis, BroermanMasterThesis, Broerman2017}. However, we cannot directly mimic the process shown in the papers to our detector for two reasons. (a) The diameter and height of our detector are only 10 cm, while the DEAP detector is a sphere with a diameter of 1.7-meter, so the inner surface area to be coated is a factor of $\sim$300 greater than ours. Assuming the same coating thickness, the TPB mass consumed in two detectors should have the same variance; consequently, the dimension of the crucible and the source (containing the crucible) should be redesigned to adapt our much smaller detector. For more information on the source, please refer to section \ref{secCoatingSubAppPro} below. (b) The DEAP-3600 detector has the ``neck-pipe'' structure through which LAr can be filled into the central detector; the auxiliary coating parts, such as cables and the source,  can go through the neck-pipe. While in our detector, we do not have the same design. During coating, some TPB molecules will deposit on the auxiliary parts instead of the desired top base surface, leaving a much thinner TPB layer area on the base. So, we must redesign a new coating process to adapt to our detector. Our solution is to split the coating job into two steps; the bottom base and the curved surface will be coated on the real detector in the first step, and the top base will be coated in the second step on a mock detector; as will be addressed in more detail in the section of~\ref{secCoatingSubTwoSteps}. 

\subsection{The coating principle and the source}
\label{secCoatingSubAppPro}

\begin{figure}[htbp]
\centering 
\includegraphics[width=.85\textwidth]{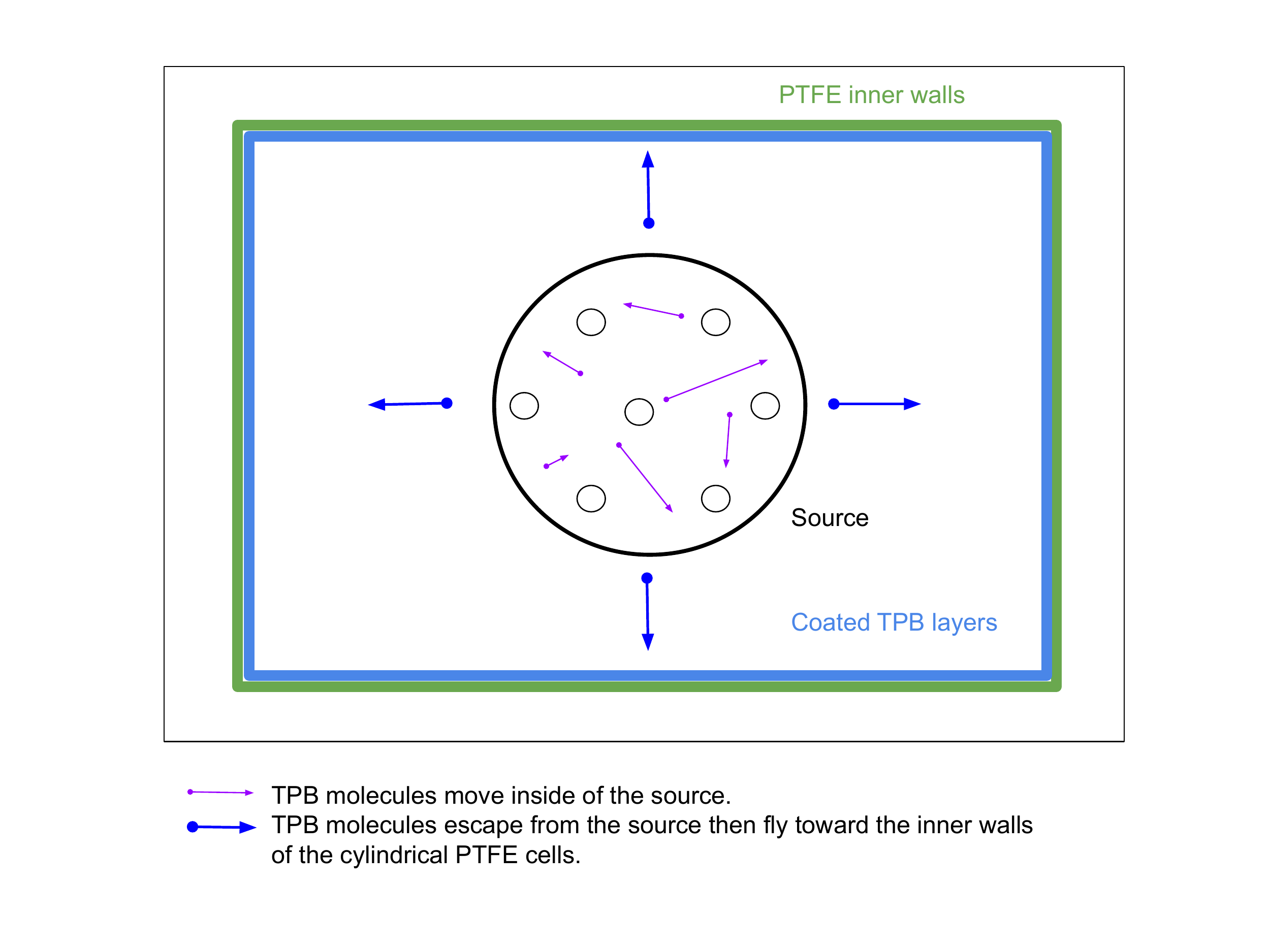}
\caption{\label{figCoatingPrinciple} A schematic drawing shows the TPB coating principle. In the center of the PTFE detector is the spherical source, inside which is a TPB powder-contained crucible (please refer to Fig.~\ref{figCoatingSource.b}). The TPB powder will sublimate into gas molecules as long as being heated at 208 $^{\circ} \mathrm{C}$~\cite{Broerman2017}. The molecules, as the purple arrows show, would scatter each other enough times; then escape randomly through a hole (as the small circles shown in the picture) on the surface of the source (as the blue arrows show). Once the molecules arrive at the colder surfaces of the PTFE, as the dark green rectangular shows, they lose kinematic energy and ultimately deposit on the surfaces, as the dark blue rectangular. The plot is not to scale.}
\end{figure}

The coating principle is schematically shown in Fig.\ref{figCoatingPrinciple}. The sphere lying in the center of the PTFE cell is the source on which tens of holes are evenly distributed; inside the sphere is a crucible containing TPB powder, which will vaporize into gas molecules as long as being heated enough. The tricky of the coating process is that it must make TPB molecules move slowly enough to ensure they scatter each other with sufficient times ($\ge$ 10 times, for instance) inside the source before randomly finding a hole to escape. This way, as a whole, the molecules would come out of the sphere isotropically, which guarantees the thickness of the coated TPB layers are nearly the same since the height and diameter of the PTFE cell are both 10 cm. When the TPB molecules arrive at the colder inner walls, they lose kinematic energy and eventually deposit on the surfaces. 

\begin{figure}	
	\centering
	\begin{subfigure}[t]{2.6in}
		\centering
		\includegraphics[scale=0.4]{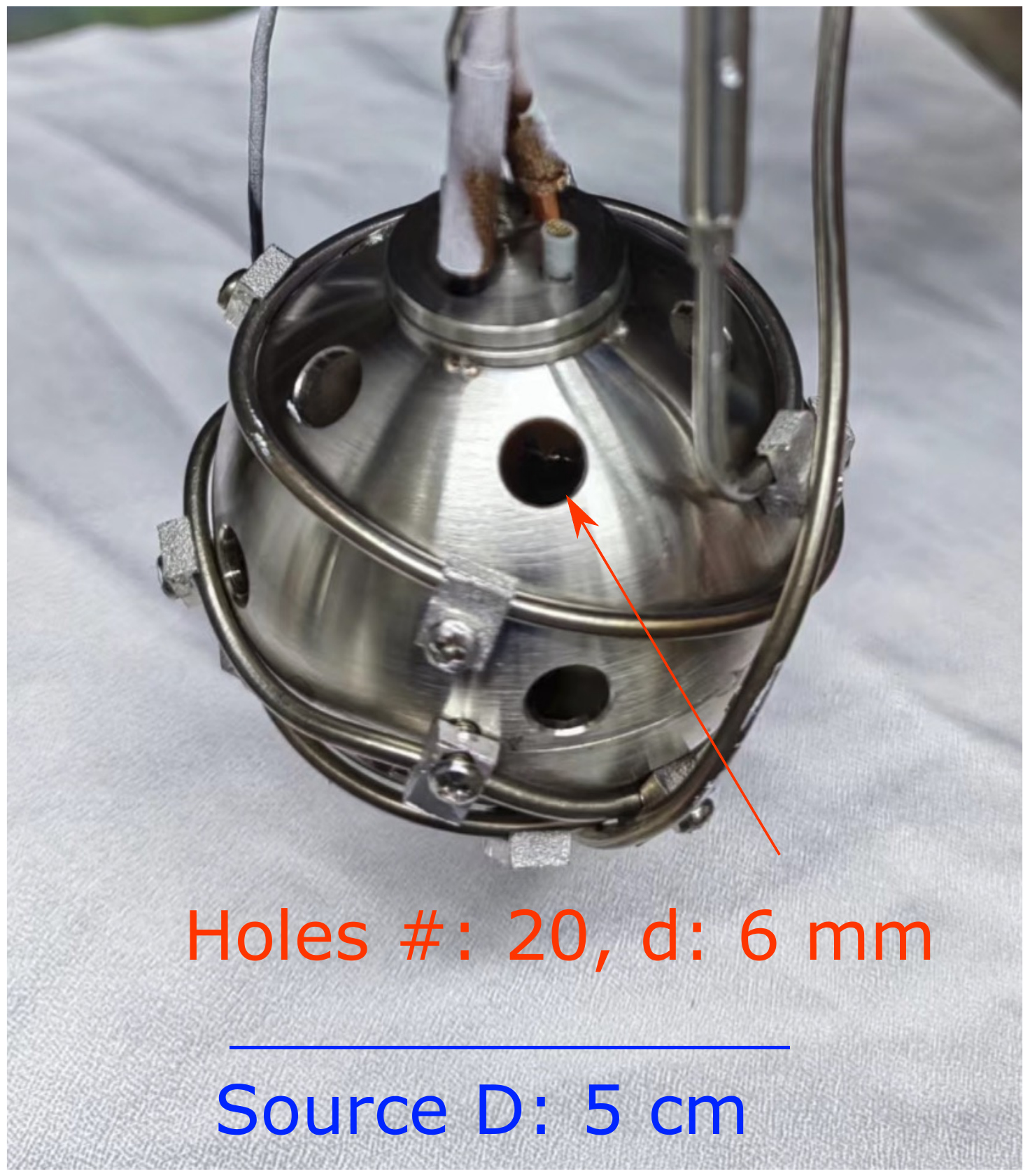}
		\caption{Fig.~\ref{figCoatingSource.a}. The source is a $\phi$5 cm sphere with 20 $\phi$6 mm holes evenly distributed on the surface. The sphere is 2 mm thick and made of stainless. The heating cable is Nimonic alloys.} \label{figCoatingSource.a}	
	\end{subfigure}
	\quad
	\begin{subfigure}[t]{2.6in}
		\centering
		\includegraphics[scale=0.48]{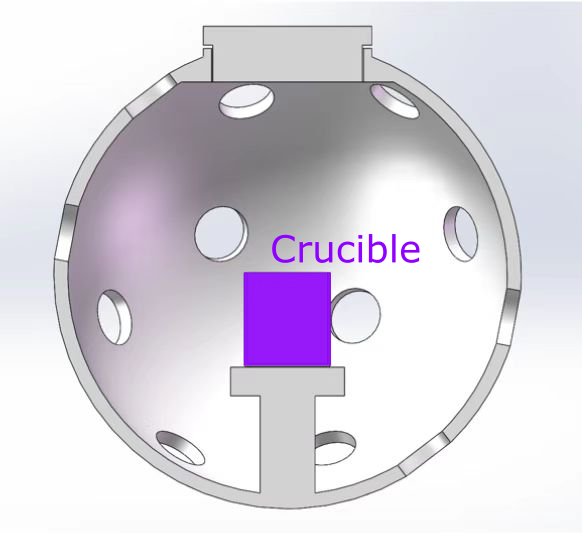}
		\caption{Fig.~\ref{figCoatingSource.b}. The schematic drawing of the source. Inside the source is a crucible that contains TPB powder.} \label{figCoatingSource.b}
	\end{subfigure}
	\caption{The picture and schematic drawing of the source we designed and built at CIAE in Beijing, China.}\label{figCoatingSource}
\end{figure}

The air pressure inside and outside the source should always have a  constant discrepancy to allow gas molecules to flow outwards stably. The crucible's heating power, the temperature of TPB molecules, and the number and size of the holes on the source are all relevant parameters to achieve the goal. 
The source's dimensions also determines whether TPB molecules can come out in an isotropic way. 
By referring to the calculations in the two papers~\cite{PollmannPhDThesis, BroermanMasterThesis}, we designed and built a 5-cm diameter source with twenty 6-mm diameters holes evenly distributed on the surface, as Fig.~\ref{figCoatingSource.a} shows. The sphere is made of stainless. We use the Nimonic alloy cable to heat the source and the crucible inside, as shown in Fig.~\ref{figCoatingSource.b}. The cable is tightly surrounded on the source's surface with screws. Heating the Nimonic alloys will eventually raise the temperature of the TPB, which is inside the crucible.

\subsection{Coating the inner walls of the cylindrical detector in two steps} 
\label{secCoatingSubTwoSteps}

\begin{figure}	
	\centering
	\begin{subfigure}[t]{2.6in}
		\centering
		\includegraphics[scale=0.465]{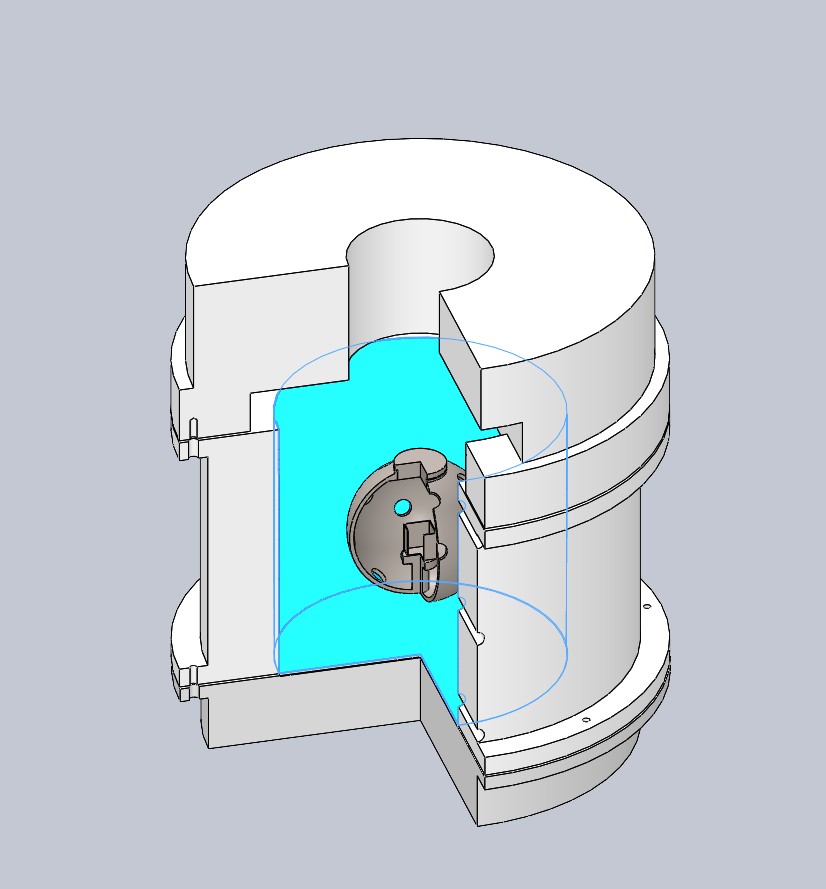}
		\caption{Fig.~\ref{figCoatingTwoSteps.a}. In the first step, we coat on the real detector. Only the bottom base and the curved surface are useful, while the top base is useless since its center will have a thinner TPB layer.} \label{figCoatingTwoSteps.a}	
	\end{subfigure}
	\quad
	\begin{subfigure}[t]{2.6in}
		\centering
		\includegraphics[scale=0.45]{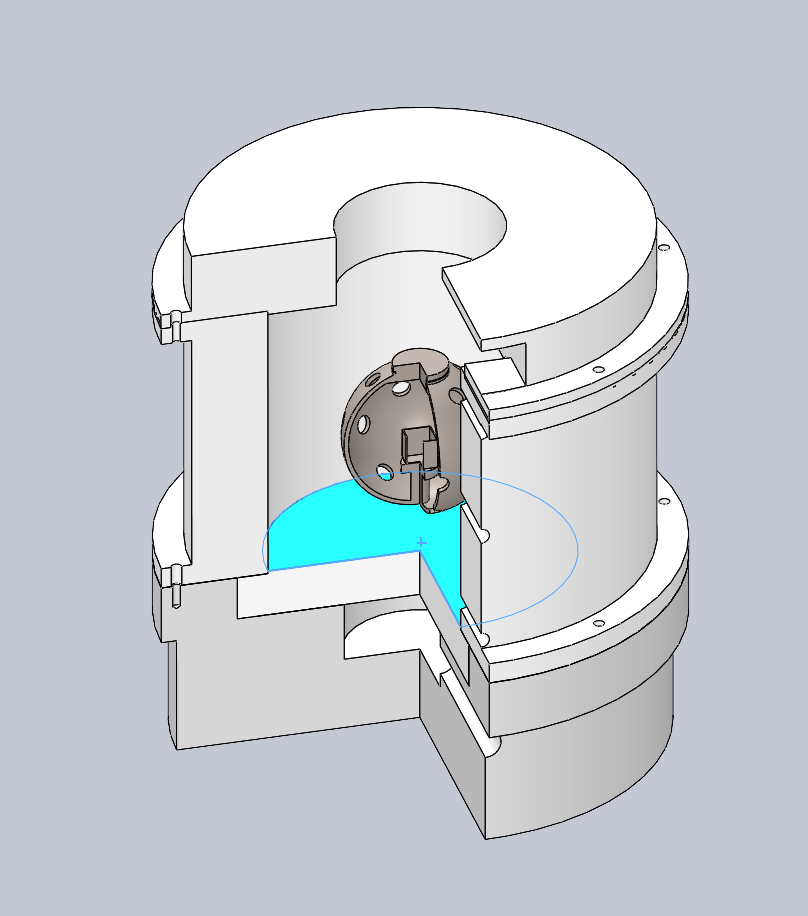}
		\caption{Fig.~\ref{figCoatingTwoSteps.b}. The second time, we coat on a mock detector (the same size as the real one). After coating, the bottom base of the mock detector can be used as the top base of the real detector.} \label{figCoatingTwoSteps.b}
	\end{subfigure}
	\caption{Schematic drawings show the procedure of coating all of the inner walls of a 10-cm cylindrical PTFE detector. In the first step, we coat on the real detector. Only the bottom base and the curved surface are useful, while the top base is useless since its center will have a thinner TPB layer. The second time, we coat on a mock detector (the same size as the real one). We use the same amount of TPB and the same coating process for the two times of coating. So the bottom base of the mock detector would have the same thickness as the inner surfaces of the real one, therefore, can be used as the top base of the real detector.}\label{figCoatingTwoSteps}
\end{figure}

Since the cables used for heating and temperature sensors must go through the top base down to the central position of the PTFE cell to hang the source, a fraction of TPB molecules will deposit on the cables during coating. Consequently, the coated layer's thickness in the center of the top base will be thinner than other areas. To have the same thickness for all inner surfaces, we developed a coating method called ``coating in two steps''. The idea is to split the whole coating process into two steps. In the first step, we coat on the real detector; only the bottom base and the curved surface will be useful since these two surfaces would have the desired thickness, as shown in Fig.~\ref{figCoatingTwoSteps.a}. As mentioned above, the top base of the real detector can not be used since the TPB layer in its center is thinner. In the second step, we coated on a same-sized mock detector. Since the two detectors have the same structure, we can take the bottom base of the mock detector out and assembly it in the real detector. Fig.~\ref{figCoatingTwoSteps.b} shows the useful bottom base being coated on a mock detector. As long as the consumed TPB mass and the coating processes are the same in the two steps, the thickness of coated TPB would be the same for the two bases and the curved surface. This way, the whole inner walls of the real detector would have a unique thick TPB layer.

\subsection{Coating thickness measurements}
\label{secCoatingThickness}
We developed three independent methods to measure the thickness of the coated TPB layer.

(i) A real-time monitoring system shows that the TPB thickness increases during coating.

(ii) Comparing the mass of a few aluminum sample films inside the detector before and after coating, then figure out the thickness of TPB being coated on the sample films.

(iii) Calculating the thickness by the consumed TPB mass and the whole area of the inner surfaces of the PTFE cell. 

We obtained consistent thickness with the three methods when coating the 10-cm PTFE detector with 0.2 g TPB. Specifically, the real-time monitoring system showed the thickness increases gradually and stably and ended up at 4.08 $\mu$m, with a 0.5\% uncertainty. The mass difference of aluminum sample films at the bottom of the 10-cm cell gave 3.93 $\pm$ 0.03 $\mu$m before and after coating. For the films in the middle of the curved surface, the difference was 4.03 $\pm$ 0.16 $\mu$m. The calculated thickness with the consumed TPB method returned 3.96 $\pm$ 0.00 $\mu$m. The results are summarized in table~\ref{tabThicknessTPB}. 

\begin{figure}[!t]	 
	\centering
        \includegraphics[width=.5\textwidth, angle = 0]{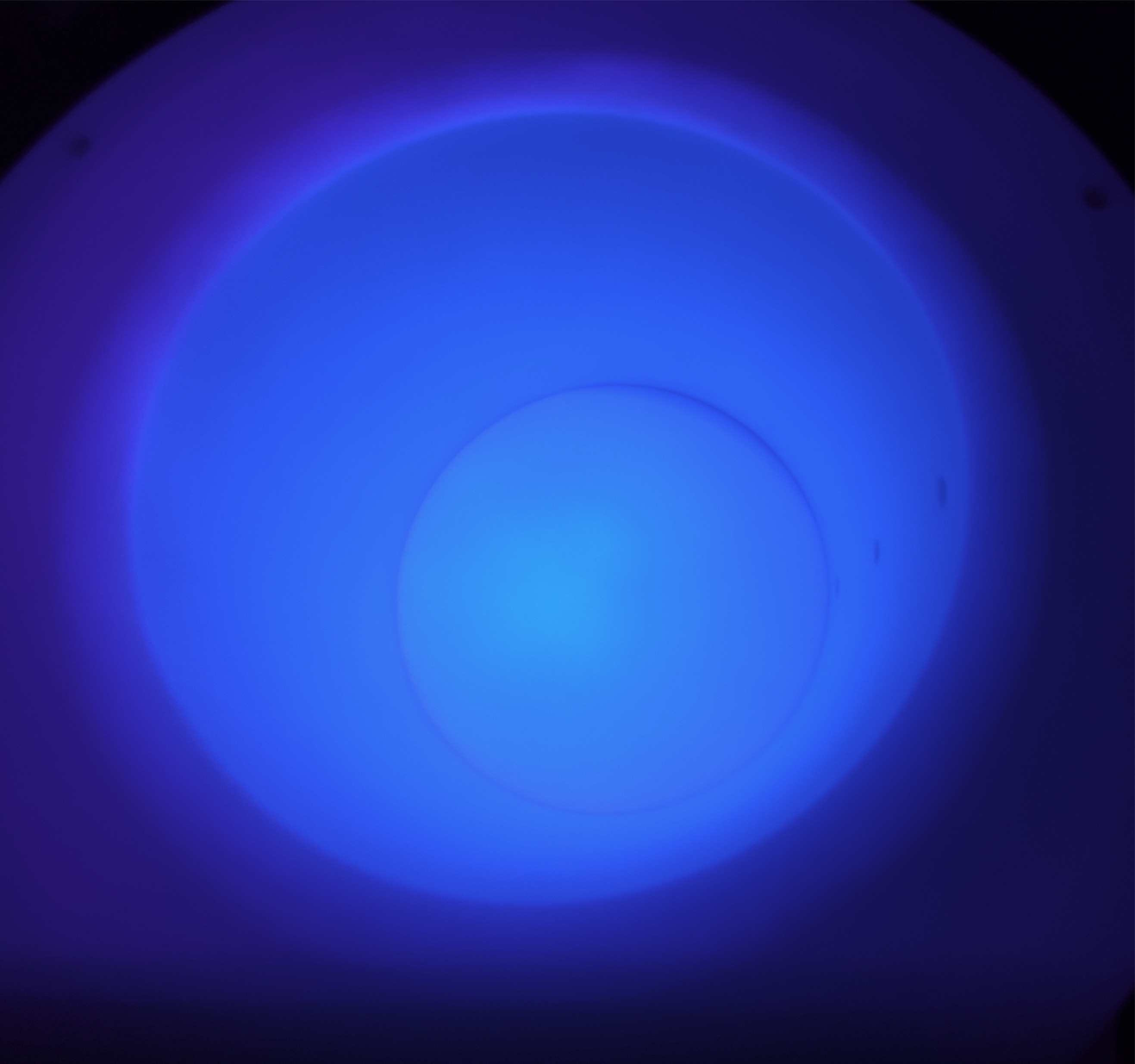}
	\caption{$\sim 3.0~ \mu$m TPB has been coated on the inner wall of a 10-cm PTFE detector. This plot is a top view being lighted up with a 395 nm wavelength torch.}\label{fig_TPBCoatdingInsideView}
\end{figure}

We verified the consistence with 0.15 g TPB mass, which results $\sim$ 3.0 $\mu$m TPB layer. Fig.~\ref{fig_TPBCoatdingInsideView} shows the 10-cm PTFE detector coated with $\sim 3.0 ~\mu$m TPB. By changing the TPB mass in the crucible, we can, in principle, coat other thickness we need.

\newcommand{\otoprule}{\midrule[\heavyrulewidth]} 
\begin{table}[ht]
   \centering
   \caption{The thickness of TPB layers obtained with three different methods.} \label{tabThicknessTPB}   
      \begin{tabular}{c c c}
      \toprule%
         \multicolumn{1}{c}{\bfseries{ Real-time monitoring } }  &
         \multicolumn{1}{c}{\bfseries{ Sample films mass } }  &
         \multicolumn{1}{c}{\bfseries{ Total consumed TPB mass} }     \\%
	  \otoprule%
4.08 $\pm$ 0.02 $\mu$m &\vtop{\hbox {\strut{3.93 $\pm$ 0.03 $\mu$m (bottom base)}} \hbox{\strut {4.03 $\pm$ 0.16 $\mu$m (curved surface) } } }       & 3.96 $\pm$ 0.00 $\mu$m\\
      \bottomrule
    \end{tabular}
\end{table}

\section{Scanning TPB coating sample films with an SEM machine}
\label{secCoatingScanning}
\begin{figure}[!t]	 
	\centering
        \includegraphics[width=.5\textwidth, angle = 0]{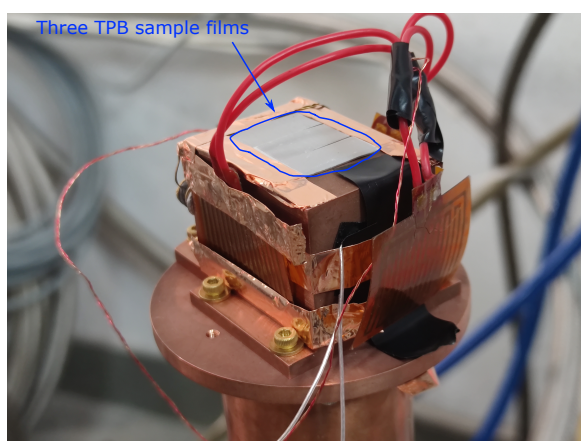}
	\caption{Three TPB sample films were mounted on the top of a copper dark box, which is inside of a GM cryocooler. The temperature sensor on the side of the dark box (invisible in the picture) shows the lowest temperature was 4.5 K.}\label{fig_ThreeTPBFilmsInsideCryocooler}
\end{figure}

To know the quality of the coating surface, we put a few alumina sample films on the curved surface and bottom base of the PTFE cell. After coating, we split the films into three groups; each group has three to four films. The films in group 1 were immersed in Liquid Nitrogen (LN) for 40 hours, then taken out to warm up naturally to Room Temperatures (RT). The films in group 2 were put in a GM cryocooler as shown in~Fig.~\ref{fig_ThreeTPBFilmsInsideCryocooler}. They had gone through three hours at 4.5 K temperature, six hours of cooling from RT to 4.5 K, and ten hours of warming up inside the cryocooler. The group 3 films have no cryogenic experience. As summarized in table~\ref{tabCryogenicTPB}. 

\begin{table}[ht]
   \centering
   \caption{The cryogenic experience of TPB films used for SEM scanning.} \label{tabCryogenicTPB}   
      \begin{tabular}{c c c}
      \toprule%
         \multicolumn{1}{c}{\bfseries{ Group 1 } }  &
         \multicolumn{1}{c}{\bfseries{ Group 2 } }  &
         \multicolumn{1}{c}{\bfseries{ Group 3 } }     \\%
	  \otoprule%
40 h in LN &	3 h at 4.5 K, 6 h cooling, 10 h warming up  &No cryogenic experience\\
      \bottomrule
    \end{tabular}
\end{table}

\begin{figure}	
	\centering
	\begin{subfigure}[t]{2.6in}
		\centering
		\includegraphics[scale=0.38]{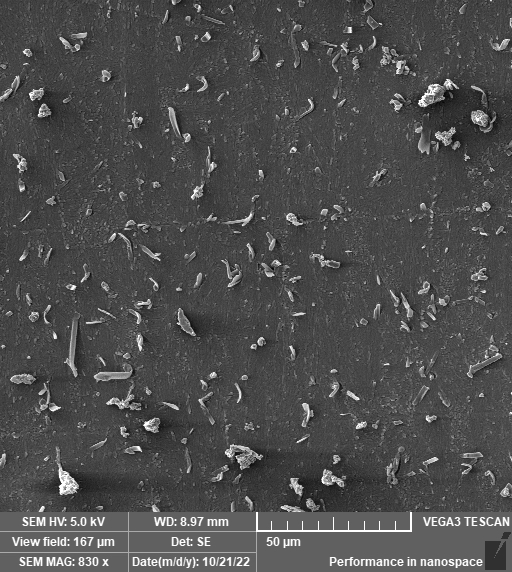}
		\caption{Fig.~\ref{figTPBCoatingFilms4KTwoResolutions.a}. SEM scanning with 50 $\mu$m resolution.} \label{figTPBCoatingFilms4KTwoResolutions.a}	
	\end{subfigure}
	\quad
	\begin{subfigure}[t]{2.6in}
		\centering
		\includegraphics[scale=0.38]{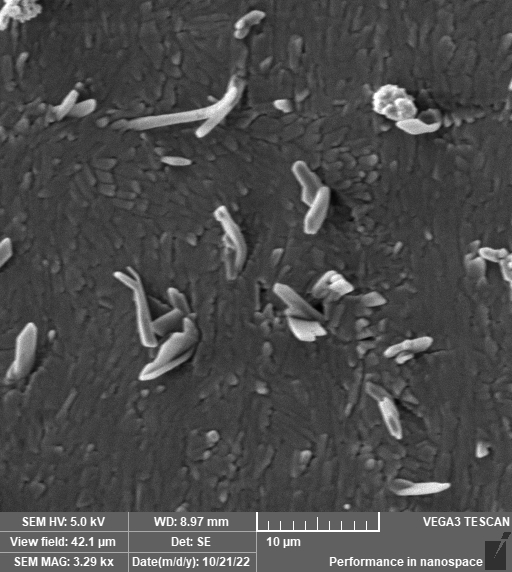}
		\caption{Fig.~\ref{figTPBCoatingFilms4KTwoResolutions.b}. SEM scanning with 10 $\mu$m resolution.} \label{figTPBCoatingFilms4KTwoResolutions.b}
	\end{subfigure}
	\caption{SEM scanning with 50 $\mu$m (Fig.~\ref{figTPBCoatingFilms4KTwoResolutions.a}) and 10 $\mu$m (Fig.~\ref{figTPBCoatingFilms4KTwoResolutions.b}) resolutions on the TPB-coated sample films went through three hours of cooling at 4 K.}\label{figTPBCoatingFilms4KTwoResolutions}
\end{figure}

\begin{figure}	
	\centering
	\begin{subfigure}[t]{2.6in}
		\centering
		\includegraphics[scale=0.38]{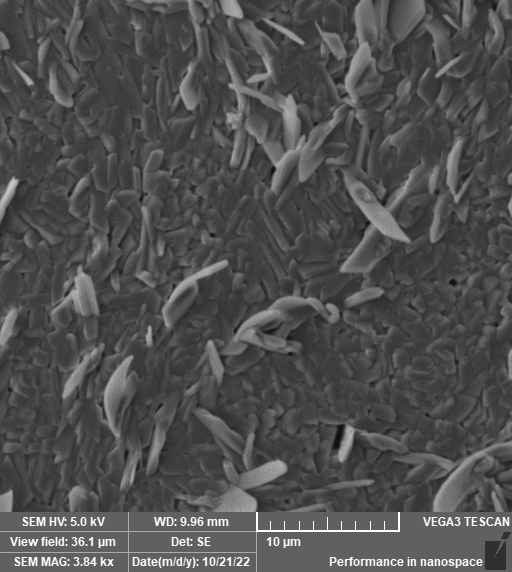}
		\caption{Fig.~\ref{figTPBCoatingFilmsLNRT10um.a}. The film had been immersed in LN for forty hours, then warmed up to RT.} \label{figTPBCoatingFilmsLNRT10um.a}	
	\end{subfigure}
	\quad
	\begin{subfigure}[t]{2.6in}
		\centering
		\includegraphics[scale=0.38]{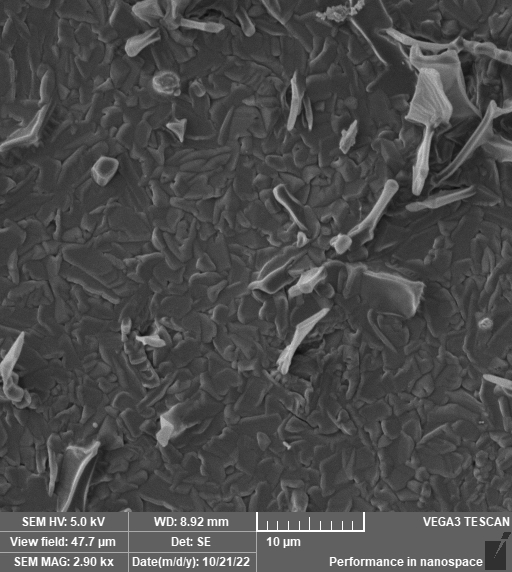}
		\caption{Fig.~\ref{figTPBCoatingFilmsLNRT10um.b}. The film has always been RT after TPB coating.} \label{figTPBCoatingFilmsLNRT10um.b}
	\end{subfigure}
	\caption{SEM scanning with 10 $\mu$m resolution on the TPB coated sample films. The film in Fig.~\ref{figTPBCoatingFilmsLNRT10um.a} had been immersed in LN for forty hours; the film Fig.~\ref{figTPBCoatingFilmsLNRT10um.b} has always been at RT.}\label{figTPBCoatingFilmsLNRT10um}
\end{figure}

We scanned these films with an SEM (Scanning Electron Microscope) machine, Tescan VEGA3~\cite{TescanVega3Website}. We scanned the sample films with two resolutions: 50 and 10 $\mu$m. The SEM images of the sample films have exposed 3 hours at 4.5 K are shown in Fig.~\ref{figTPBCoatingFilms4KTwoResolutions.a}  and Fig.~\ref{figTPBCoatingFilms4KTwoResolutions.b}, with a resolution of 50 and 10 $\mu$m, respectively. Fig.~\ref{figTPBCoatingFilmsLNRT10um.a} shows the SEM image scanned with a resolution of 10 $\mu$m for the sample film immersed forty hours in LN. Fig.~\ref{figTPBCoatingFilmsLNRT10um.b} is the film has no any cryogenic experience.

After studying the images as shown in Fig.~\ref{figTPBCoatingFilms4KTwoResolutions.b}, Fig.~\ref{figTPBCoatingFilmsLNRT10um.a}, and Fig.~\ref{figTPBCoatingFilmsLNRT10um.b}, we found: (a) TPB has been well coated on all of the sample films, (b) There exist needle-like structures shown in the SEM images which are TPB molecules formed crystals, (c) comparing these scanned images, we can not find any noticeable difference. 

\section{Summary}
\label{secSummary}

ALETHEIA is a promising project in hunting for low-mass DM. In this paper, we developed the process of coating $\sim \mu$m TPB on the inner walls of a cylindrical detector. Further, the SEM scanning images show (a) TPB has been well coated on the sample films, and (b) the films being exposed to 4.5 K do not show any noticeable difference comparing to the ones at RT and LN temperature. This work thus gives us confidence to implement TPB as the wavelength of our LHe TPCs.

Next, we will study the TPB conversion efficiency at LHe temperature, though the efficiency has been fully studied at room temperature, as shown in reference~\cite{Benson18}. 

\acknowledgments

This work was supported by the Yuanzhang funding of CIAE. We thank our colleagues at CIAE, Dr. Yunju Li and Miss Chen Chen for their help on TPB coating, Dr. Ping Fan and Miss Xinxin Li for supporting on SEM machine operation.

\bibliography{TPBCoatingLowTempTests}

\providecommand{\noopsort}[1]{}\providecommand{\singleletter}[1]{#1}

\providecommand{\href}[2]{#2}\begingroup\raggedright\begin{thebibliography}{10}

\bibitem{DarkSideWebsite}
\emph{{DarkSide official website at LNGS}},  Nov, 2022.

\bibitem{DEAP3600Website}
\emph{{DEAP3600 official website}},  Nov, 2022.

\bibitem{LUXWebsite}
\emph{{LUX official website}},  Nov, 2022.

\bibitem{LZWebsite}
\emph{{LZ official website}},  Nov, 2022.

\bibitem{PandaXWebsite}
\emph{{PandaX official website}},  Nov, 2022.

\bibitem{XENONWebsite}
\emph{{XENON official website}},  Nov, 2022.

\bibitem{heliumAbundance}
J.~Emsley, \emph{{Nature's Building Blocks: An A-Z Guide to the Elements}},
  Oxford University Press, 2nd~ed. (Oct, 2011).

\bibitem{JEWELL1979682}
C.~Jewell and P.~McClintock, \emph{A note on the purity of liquid helium-4},
  {\emph{Cryogenics} {\bfseries 19} (1979) 682}.

\bibitem{ALETHEIA-EPJ-22}
J.~Liao, Y.~Gao, Z.~Jiang, Z.~Liang, Z.~OuYang, Z.~Peng et~al., ``{ALETHEIA}:
  {H}unting for {L}ow-mass {D}ark {M}atter with {L}iquid {H}elium {TPC}s.''
  Submitted to {EPJP}, Sep, 2022.
\newblock https://arxiv.org/abs/2209.02320.

\bibitem{ALETHEIA_2019Review}
R.~Gaitskell, D.~Hooper, J.~Liu, D.~McKinsey, G.~Seidel and I.~Takeyaso,
  ``{ALHET Draft Report, review at Peking University, 191016 (Report Draft
  191028 v6)}.'' Private document (No publication), Oct, 2019.

\bibitem{DMWS2019PKU}
J.~Liao and Q.~Cao, ``{Dark matter (WIMPs) direct detection workshop}.''
  https://indico.ihep.ac.cn/event/10369/overview, Oct, 2019.

\bibitem{McKinsey03}
D.N.~McKinsey, C.R.~Brome, S.N.~Dzhosyuk, R.~Golub, K.~Habicht, P.R.~Huffman
  et~al., \emph{Time dependence of liquid-helium fluorescence},
  \href{https://doi.org/10.1103/PhysRevA.67.062716}{\emph{Phys. Rev. A}
  {\bfseries 67} (2003) 062716}.

\bibitem{Ito2012}
T.M.~Ito, S.M.~Clayton, J.~Ramsey, M.~Karcz, C.-Y.~Liu, J.C.~Long et~al.,
  \emph{Effect of an electric field on superfluid helium scintillation produced
  by $\ensuremath{\alpha}$-particle sources},
  \href{https://doi.org/10.1103/PhysRevA.85.042718}{\emph{Phys. Rev. A}
  {\bfseries 85} (2012) 042718}.

\bibitem{ItoSeidel13}
T.M.~Ito and G.M.~Seidel, \emph{Scintillation of liquid helium for low-energy
  nuclear recoils},
  \href{https://doi.org/10.1103/PhysRevC.88.025805}{\emph{Phys. Rev. C}
  {\bfseries 88} (2013) 025805}.

\bibitem{Seidel14}
G.M.~Seidel, T.M.~Ito, A.~Ghosh and B.~Sethumadhavan, \emph{Charge distribution
  about an ionizing electron track in liquid helium},
  \href{https://doi.org/10.1103/PhysRevC.89.025808}{\emph{Phys. Rev. C}
  {\bfseries 89} (2014) 025808}.

\bibitem{Ito16}
T.M.~Ito, J.C.~Ramsey, W.~Yao, D.H.~Beck, V.~Cianciolo, S.M.~Clayton et~al.,
  \emph{An apparatus for studying electrical breakdown in liquid helium at 0.4
  k and testing electrode materials for the neutron electric dipole moment
  experiment at the spallation neutron source},
  \href{https://doi.org/10.1063/1.4946896}{\emph{Review of Scientific
  Instruments} {\bfseries 87} (2016) 045113}
  [\href{https://arxiv.org/abs/https://doi.org/10.1063/1.4946896}{{\ttfamily
  https://doi.org/10.1063/1.4946896}}].

\bibitem{Phan20}
N.S.~Phan, V.~Cianciolo, S.M.~Clayton, S.A.~Currie, R.~Dipert, T.M.~Ito et~al.,
  \emph{Effect of an electric field on liquid helium scintillation produced by
  fast electrons},
  \href{https://doi.org/10.1103/PhysRevC.102.035503}{\emph{Phys. Rev. C}
  {\bfseries 102} (2020) 035503}.

\bibitem{SpiceHeRald21}
S.A.~Hertel, A.~Biekert, J.~Lin, V.~Velan and D.N.~McKinsey, \emph{Direct
  detection of sub-gev dark matter using a superfluid $^{4}\mathrm{He}$
  target},
  \href{https://doi.org/https://link.aps.org/doi/10.1103/PhysRevD.100.092007}{\emph{Phys.
  Rev. D} {\bfseries 100} (2019) 092007}.

\bibitem{Benson18}
C.~Benson, G.D.~Orebi~Gann and V.~Gehman, \emph{Measurements of the intrinsic
  quantum efficiency and absorption length of tetraphenyl butadiene thin films
  in the vacuum ultraviolet regime},
  \href{https://doi.org/10.1140/epjc/s10052-018-5807-z}{\emph{The European
  Physical Journal C} {\bfseries 78} (2018) }.

\bibitem{Broerman2017}
B.~Broerman, M.~Boulay, B.~Cai, D.~Cranshaw, K.~Dering, S.~Florian et~al.,
  \emph{Application of the {TPB} wavelength shifter to the {DEAP}-3600
  spherical acrylic vessel inner surface},
  \href{https://doi.org/10.1088/1748-0221/12/04/p04017}{\emph{Journal of
  Instrumentation} {\bfseries 12} (2017) P04017}.

\bibitem{DarkSide20k17}
C.E.~Aalseth, F.~Acerbi, P.~Agnes, I.F.M.~Albuquerque, T.~Alexander, A.~Alici
  et~al., \emph{Darkside-20k: A 20 tonne two-phase lar tpc for direct dark
  matter detection at lngs},
  \href{https://doi.org/10.1140/epjp/i2018-11973-4}{\emph{The European Physical
  Journal Plus} {\bfseries 133} (2018) }.

\bibitem{AGNES2015456}
P.~Agnes, T.~Alexander, A.~Alton, K.~Arisaka, H.~Back, B.~Baldin et~al.,
  \emph{First results from the darkside-50 dark matter experiment at laboratori
  nazionali del gran sasso},
  \href{https://doi.org/https://doi.org/10.1016/j.physletb.2015.03.012}{\emph{Physics
  Letters B} {\bfseries 743} (2015) 456}.

\bibitem{PollmannPhDThesis}
T.~Pollmann, \emph{Alpha backgrounds in the DEAP Dark Matter search
  experiment}, {PhD} dissertation, Queen's University, Department of Physics,
  Engineering physics and Astronomy, 2012.

\bibitem{BroermanMasterThesis}
B.~Broerman, \emph{On the Development of the Wavelength Shifter Deposition
  System for the DEAP-3600 Dark Matter Search Experiment}, {Master}
  dissertation, Queen's University, Department of Physics, Engineering physics
  and Astronomy, 2015.

\bibitem{TescanVega3Website}
\emph{https://www.tescan.com/tescan-vega3-an-analytical-tool-for-life-sciences/},
  Oct, 2022.

\end{thebibliography}\endgroup


\end{document}